\documentstyle[11pt,aaspp4]{article}
\singlespace

\def\Kp{\ifmmode{K^{\prime}}\else{${K^{\prime}}$}\fi}
\def\MKp{\ifmmode{\overline M_{K^{\prime}}}\else{$\overline M_{K^{\prime}}$}\fi}
\def\mKp{\ifmmode{\overline m_{K^{\prime}}}\else{$\overline m_{K^{\prime}}$}\fi}
\def\MK{\ifmmode{\overline M_{K}}\else{$\overline M_{K}$}\fi}
\def\mK{\ifmmode{\overline m_{K}}\else{$\overline m_{K}$}\fi}
\def\MI{\ifmmode{\overline M_{I}}\else{$\overline M_{I}$}\fi}
\def\mI{\ifmmode{\overline m_{I}}\else{$\overline m_{I}$}\fi}
\def\MH{\ifmmode{\overline M_{H}}\else{$\overline M_{H}$}\fi}
\def\mH{\ifmmode{\overline m_{H}}\else{$\overline m_{H}$}\fi}
\def\MJ{\ifmmode{\overline M_{J}}\else{$\overline M_{J}$}\fi}
\def\mJ{\ifmmode{\overline m_{J}}\else{$\overline m_{J}$}\fi}
\def\p0/p1{\ifmmode{P_0/P_1}\else{$P_0/P_1$}\fi}
\def\snsbf{\ifmmode{\xi}\else{$\xi$}\fi}
\def\E{\ifmmode{E(k)}\else{$E(k)$}\fi}
\def\mg2{Mg$_2$}
\def\Mg2{Mg$_2$}
\def\v-i{\ifmmode{(V{-}I)}\else{$(V{-}I)$}\fi}
\def\m-M{\ifmmode{(m{-}M)}\else{$(m{-}M)$}\fi}
\def\m-Mbar{\ifmmode{(\overline m{-}\overline M)}\else{$(\overline m{-}\overline M)$}\fi}
\def\kms{\ifmmode{{\rm km\,s}^{-1}}\else{km\,s$^{-1}$}\fi}
\def\kmsmpc{km\,s$^{-1}$\,Mpc$^{-1}$}
\def\eps{\ifmmode{{\rm e}^-\,{\rm s}^{-1}}\else{e$^-$\,s$^{-1}$}\fi}
\def\epspp{\ifmmode{{\rm e}^-\,{\rm s}^{-1}\,{\rm pixel}^{-1}}\else{e$^-$\,s$^{-1}$\,pixel$^{-1}$}\fi}

\begin{document}

\title{The Infrared Surface Brightness Fluctuation Distances\\
to the Hydra and Coma Clusters}

\author{Joseph B. Jensen\altaffilmark{1}}
\author{John L. Tonry}
\and
\author{Gerard A. Luppino}
\affil{Institute for Astronomy, University of Hawaii\\
	2680 Woodlawn Drive, Honolulu, HI  96822\\
	e-mail: jjensen@gemini.edu, jt@avidya.ifa.hawaii.edu, 
	ger@hokupa.ifa.hawaii.edu} 
\authoraddr{2680 Woodlawn Drive, Honolulu, HI  96822}
\altaffiltext{1}{Currently with the Gemini 8-m Telescopes Project,
        180 Kinoole St. Suite 207, Hilo, HI  96720.  The Gemini 8-m 
        Telescopes Project is managed by the Association of 
        Universities for Research in Astronomy, for the National Science 
        Foundation and the Gemini Board, under an international 
        partnership agreement.}

\slugcomment{Submitted to {\it The Astrophysical Journal,} April 30, 1998}

\begin{abstract}

We present IR surface brightness fluctuation (SBF) distance
measurements to NGC~4889
in the Coma cluster and to NGC~3309 and NGC~3311 in the Hydra cluster.
We explicitly corrected for the contributions to the fluctuations 
from globular clusters, background galaxies, and
residual background variance.
We measured a distance of $85\,{\pm}\,10$ Mpc to NGC~4889  
and a distance of $46\,{\pm}\,5$ Mpc to the Hydra cluster.
Adopting recession velocities of
$7186\,{\pm}\,428$ \kms\ for Coma and $4054\,{\pm}\,296$ \kms\ 
for Hydra gives a mean Hubble constant of 
$H_0\,{=}\,87\,{\pm}\,11$\kms\ Mpc$^{-1}$.
Corrections for residual variances were a significant fraction of the 
SBF signal measured, and, if underestimated, would bias our measurement
towards smaller distances and larger values of $H_0$.
Both NICMOS on the Hubble Space Telescope
and large-aperture ground-based telescopes with new IR detectors
will make accurate SBF distance measurements possible to 100 Mpc and
beyond.
\end{abstract}


\section{Introduction}

Measuring accurate and reliable distances is a critical part of the 
quest to measure the Hubble constant $H_0$.  
Until recently, different techniques
for estimating extragalactic distances produced results that differed by
as much as a factor of two, while the statistical uncertainties of individual 
techniques were much smaller.
Identifying and removing systematic errors from distance measurements 
has been a difficult and lengthy process,
but progress is now being made on a number of fronts towards resolving 
discrepancies in distance measurements.
Detailed summaries of many different distance measurement
techniques are
found in the reviews by Jacoby et al. (1992) 
and in the proceedings of a recent symposium on the extragalactic distance
scale (Livio, Donahue, \& Panagia 1997).

Reliably determining $H_0$ also requires the measurement of accurate 
radial velocities.  In the local universe, peculiar velocities are larger
than the Hubble flow we must measure.  
To avoid this problem, we must accurately measure distances great enough
that local velocity perturbations are negligible.   
Recently, however, several studies have measured residual velocities on
uncomfortably large scales (Lynden-Bell et al. 1988; Mould et al. 1991; 
Mathewson, Ford, \& Buchhorn 1992; Lauer \& Postman 1994)
Large-scale motions are difficult to explain with standard models
of hierarchical structure formation, so it is crucial to our understanding
of cosmology that we confirm the reality of such large flows
with accurate distance measurements.
Large-scale motions can also bias measurements of $H_0$
so it is important that accurate distances be measured to galaxies
in many directions and to distances beyond 10,000 \kms.  

Surface brightness fluctuations (SBFs) have proven to be a remarkably accurate
distance indicator (Tonry et al. 1997).  
Ground-based optical SBFs reach ${\sim}3000$ \kms; 
IR SBFs are some 30 times brighter than optical fluctuations, and can 
potentially reach much greater distances.  
We explored the advantages of measuring
SBFs in the \Kp\ band (2.1 \micron) and calibrated the \Kp\ SBF distance
scale (Jensen, Tonry, \& Luppino 1998, hereafter JTL; 
Jensen, Luppino, \& Tonry 1996).  
From these studies we learned that it is important to account for the
contributions to the IR SBFs from globular clusters and background 
galaxies.  JTL and Jensen et al. (1996) also 
concluded that residual noise in IR images
that is correlated from pixel to pixel can dominate SBFs 
when the ratio of signal to noise ($S/N$) is small.

This paper presents measurements of the \Kp\ SBF 
distances to NGC~3309 and NGC~3311 in the Hydra cluster and to
NGC~4889 in the Coma cluster to establish IR SBF
as a technique useful to ${\sim}7000$ \kms.
This is the first step towards an accurate measurement of the
IR SBF Hubble constant and a better determination of peculiar velocities
on scales of ${\sim}5000$ \kms.
At 7000 \kms, the peculiar velocity of the Coma cluster should be less
than 10\% of its Hubble velocity.  
At half the distance to Coma, 
SBF measurements of Hydra cluster galaxies can be used to
test our techniques and
to confirm the reliability of the more distant Coma measurements.
Tully-Fisher and $D_n{-}\sigma$ distance measurements both show Hydra
to be nearly at rest with respect to the Hubble flow
(Han \& Mould 1992; Faber et al. 1989),
so we can use the distance to Hydra to confirm the Coma cluster determination
of $H_0$.

\section{Observations and Data Reduction}

The data described in this paper were obtained using
the University of
Hawaii 2.24~m telescope on 1996 March 1 and April 25--28.
We used the new science-grade ``HAWAII'' infrared array, a 1024$^2$-pixel
HgCdTe detector mounted in the QUIRC camera (Hodapp et al. 1996).  
At f/10, QUIRC has a pixel scale of 0\farcs1886~pixel$^{-1}$, 
producing a field of view 3\farcm2 on a side.  
We observed using the University of Hawaii \Kp\ filter centered at 
2.1 \micron, 
which is much less sensitive to the thermal background than
the standard $K$ filter (Wainscoat \& Cowie 1992). 
Conditions on these nights were photometric and the seeing at \Kp\ averaged 
0\farcs75 FWHM.  
The dark current of the array was less than 1 \eps\ 
and the correlated double-sample read noise
was ${\lesssim}15$ e$^-$.  The banded sensitivity patterns and anomalous dark
current features common in earlier IR arrays have been reduced  
significantly in the new HAWAII device.  
In addition, the new array has
far fewer bad pixels (${\lesssim}1\%$) and higher quantum efficiency than 
previous IR arrays.  
The dark current in this device is locally sensitive
to bright objects, and residual images of bright or saturated objects persist
for several minutes.  
Individual integration times were kept short enough to prevent
the centers of the galaxies from approaching saturation, and sky frames
were acquired after every galaxy image.  

\begin{deluxetable}{lcc}
\tablecaption{Hydra and Coma Observational Data\label{photdata}}
\tablewidth{0pc}
\tablehead{
&\colhead{Hydra}&
\colhead{Coma}}
\startdata
Galaxies\dotfill& NGC~3309, NGC~3311 & NGC~4889 \nl
Observation dates\dotfill& 1996 Mar 1,  & 1996 Apr 25--28 \nl
                  & 1996 Apr 25, 26, 28 & \nl
$m_1$ (mag)\tablenotemark{a}\dotfill& $23.02\pm0.01$ & $22.99\pm0.01$ \nl
$\sec z$\dotfill& 1.594 & 1.287 \nl
$A_{atm}$ (${\rm mag}/\sec z$)\dotfill& 0.088 & 0.065 \nl
Seeing FWHM\dotfill& 0\farcs85 & 0\farcs75 \nl
Sky brightness (mag arcsec$^{-2}$)\dotfill& 13.62 & 13.70 \nl
Sky brightness (\epspp)\dotfill& 206 & 185 \nl
Residual sky (\epspp)\dotfill& $0.62\pm0.10$ & $0.88\pm0.05$ \nl
$t_{exp}$ (s)\dotfill& 90 & 120\nl
$t_{tot}$ (s)\dotfill& 9,090 & 15,960 \nl
$A_B$ (mag)\tablenotemark{b}\dotfill& 0.17 & 0.05 \nl
$A_{\Kp}$ (mag)\dotfill& 0.012 & 0.003 \nl
\enddata
\tablenotetext{a}{Magnitude of a source yielding 1 \eps.}
\tablenotetext{b}{Burstein \& Heiles 1984}
\end{deluxetable}

We integrated for 90 to 120 s at a time on each galaxy, 
and then moved ${\sim}400$\arcsec\ 
to sample the sky.  
Sky fields were chosen to be relatively free of galaxies and
bright stars.  
Exposure times, sky brightnesses, and other observational parameters
are listed in Table~\ref{photdata}.
UKIRT faint standard stars were observed several times during 
each night (Casali \& Hawarden 1992).  
Observing conditions were excellent during both
1996 March and April, and our photometric zero point was internally 
consistent to 0.01 mag.
We used the same standard stars and filter described in the \Kp\ SBF
calibration study (JTL)
to minimize systematic uncertainties.
The atmospheric extinction coefficient measured during the 1996 April
run was 0.065 mag~air~mass$^{-1}$ and was consistent enough 
from night to night to be used for the entire run.  
During 1996 March the extinction
coefficient was determined to be 0.115 mag~air~mass$^{-1}$.
NGC~3309 and NGC~3311 were 
observed during both runs, and the photometric data listed
in Table~\ref{photdata} are weighted mean values.

We adopted a galactic foreground extinction correction of 
$A_{\Kp}\,{=}\,0.068A_B$
based on the results of Cohen et al. (1981) and as described in 
JTL.
Extinction values in the $B$ band were taken
from Burstein \& Heiles (1984). 
Extinction corrections are small, and their 
associated uncertainties are negligible.
K corrections were computed 
for the redshifts of the Coma and Hydra clusters by 
computing $K$-band fluctuation 
magnitudes for model stellar populations at redshifts in the range
$0.0\,{\leq}\,z\,{\leq}\,0.03$.  
We used the K corrections listed in Table~6 of JTL, 
computed using
the stellar population models from Worthey (1994) and 
G. Worthey (1997, private communication) assuming a
metallicity of [Fe/H]\,=\,0.
For the Coma cluster at $z\,{=}\,0.025$, the K correction is between
0.003 and $-0.011$ mag for stellar populations with ages from 12 to 17 Gyr.
For old, metal-rich stellar populations, 
the K correction is negligible.

The first step in the image reduction procedure 
was to create an average sky image.  
We masked stars and galaxies in individual sky frames before averaging
5 to 7 images to form a sky image.  Individual exposures were 120~s,
so a stack of 7 sky frames spans about 30 minutes of observing time,
which is roughly the time scale over which variations in sky
brightness typically begin to be significant.
Next, an average sky image was subtracted from each galaxy frame prior to 
dividing by a normalized flat field image.  Dark current is subtracted
along with the sky background.  Cosmic rays were then removed from each
image.  Finally, the cleaned images were registered to the nearest pixel 
and good pixels averaged to form the final galaxy image.    
Sub-pixel registration was not used because it modifies the noise 
characteristics of the background, introducing correlations in the noise
between pixels that were uncorrelated in the individual images.  
The registered \Kp\ images are 
displayed in Figures~1 and 2.



Because the sky level changes on short time scales, sky subtraction 
left a non-zero background level that we measured and removed.
We assumed the residual sky offset was constant across the field of view
and applied a simple offset correction.  The galaxies we observed are 
large and fill most of the field of view.  To determine the residual 
sky level,
we fitted a deVaucouleurs $r^{1/4}$ profile to the surface brightness of
the galaxy measured in elliptical annuli.  We then adjusted the background
level to minimize the deviation from an $r^{1/4}$ profile.
The residual sky corrections listed in Table~\ref{photdata} were applied
before proceeding with the SBF analysis.

\section{Surface Brightness Fluctuation Analysis\label{howtosbf}}

The techniques for measuring distances using SBFs
were first laid out by Tonry \& Schneider (1988).
The procedures used to measure SBFs in this study are the same as those
presented in detail by JTL.  
We summarize only the principal steps here.
First we fitted a smooth model to the galaxy surface brightness after
the residual sky brightness had been subtracted.  
Extraneous objects were masked before fitting.  
After the galaxy had been fitted, subtracted, and compact sources masked, 
residual large-scale variations were fitted and subtracted.  
The masked residual image was then Fourier transformed and the 
two-dimensional power spectrum computed.
The power spectrum is the convolution of the stellar SBFs with 
an expectation power spectrum \E, which includes the normalized
power spectrum of the point spread function (PSF), the galaxy
surface brightness profile, and the power spectrum of the mask.
The details of the procedure are described by JTL.  

The power spectrum was fitted with the sum of
the SBF power $P_0$ multiplied by the expectation power spectrum \E\
and a white noise component $P_1$.
We measured $P_0$ in annular regions centered on the galaxy.
The contribution to the fluctuation power from undetected GCs and
background galaxies ($P_r$) was estimated and subtracted from $P_0$.
The residual spatial variance ($P_g$) remaining after subtracting the bright
background and galaxy was also estimated and subtracted from $P_0$.
The methods we used to determine $P_r$ and $P_g$ are described in detail
below
and by JTL.   
The fluctuation magnitude is defined as
\begin{equation}
\mKp = -2.5~\log(P_0{-}P_r{-}P_g)~+~m_1~-~A_{atm}\sec z~-~A_{\Kp}
\label{mkp}
\end{equation}
where $P_0$ is the SBF power in \epspp\
and $m_1$ is the magnitude of a source yielding
one \eps\ at zero air mass.

To determine the size of the contribution to the stellar SBFs from undetected 
GCs and background galaxies $P_r$,
we first measured the magnitudes of all the objects in the image.
GC and galaxy luminosity functions were fitted 
as a function of radius from the center of the galaxy 
(as described by JTL).
Objects brighter than the completeness limit were masked, and we integrated
the luminosity functions beyond the completeness cutoff 
to compute the numbers of faint unmasked GCs and galaxies 
that add variance to $P_0$.
The size of the $P_r$ correction is clearly a function of the 
limiting magnitude of the observation.  
JTL showed that even though the contrast between
stellar SBFs and GCs is a factor of ten higher in the IR than at 
optical wavelengths, the fraction of the total IR SBF power from GCs 
and galaxies can still be ${\sim}50\%$, or larger in galaxies 
like NGC~3311 that have unusually large populations of GCs.  

At \Kp, the sky background is hundreds of times brighter than in the
optical bands. 
Even when IR integrations are long enough to detect SBFs, 
the majority of the GCs remain undetected in the background noise.
When the fit to the GC and galaxy luminosity functions is derived 
from only the few brightest objects, it is poorly constrained and 
the uncertainty in $P_r$ is the principal source of uncertainty in \mKp.
JTL showed that using deep optical images to identify the locations of GCs 
and background galaxies undetected in the IR 
can reduce $P_r$ to negligible levels.
From the optical image, a mask was created and used to remove
objects from the \Kp\ image that are too faint to be detected at \Kp.
With much lower background levels, the optical image reaches a much fainter
limiting magnitude and the resulting mask removes a greater fraction of
the GCs and galaxies.
We used the composite optical/IR technique to remove 
the GC and galaxy contributions from the Coma and Hydra data.

For the observations of relatively nearby galaxies studied 
by JTL, 
the residual variance $P_r$ from GCs and galaxies after applying a mask 
generated from deep $I$-band images was less than 1\% and was ignored.  
For the Coma cluster data presented here, the optical images only reach
GCs two to three mag brighter than the peak of the luminosity function,
requiring a careful correction for the the clusters and galaxies that
remain unmasked.
For the present study, we developed a straightforward 
method for estimating $P_r$
in the \Kp\ observations based on the $I$-band photometry.
We transformed the $I$-band magnitudes to \Kp-band assuming a uniform
color for each of the GC and galaxy populations.
Values were computed for the $(I{-}K)$ colors of the 
GCs and galaxies by adopting mean $(V{-}K)$ colors from Frogel et al. (1978)
and mean \v-i colors from Ajhar \& Tonry (1994) (GCs) and 
JTL (galaxies).
We adopted $(I{-}K)\,{=}\,1.22$ for GCs and 2.1 for galaxies.
The coordinates of the objects
were scaled and translated to the frame defined by the \Kp\ image. 
We then fitted the luminosity functions, created a mask of objects
brighter than the completeness limit, and estimated the $K$-band
correction $P_r$ for our observations.

To fit the GC luminosity function, we assumed the width, form, and peak
magnitude of the GCLF, which required a preliminary estimate of the distance.
Fitting the galaxy luminosity function required a power-law slope and
normalization. 
JTL described the detailed forms of the 
luminosity functions used to estimate $P_r$.
The fluctuation magnitude is not sensitive to the fit parameters provided
that objects ${\sim}2$ mag from the GCLF peak and brighter are measured, 
fitted and masked. 
The variance is dominated by the brightest objects, so \mKp\ is insensitive
to errors in the fit to the faint end of the luminosity functions.

\section{\Kp\ SBF Magnitudes and Distances\label{coma}}

\begin{figure}
\figurenum{3}
\epsscale{0.65}
\plotone{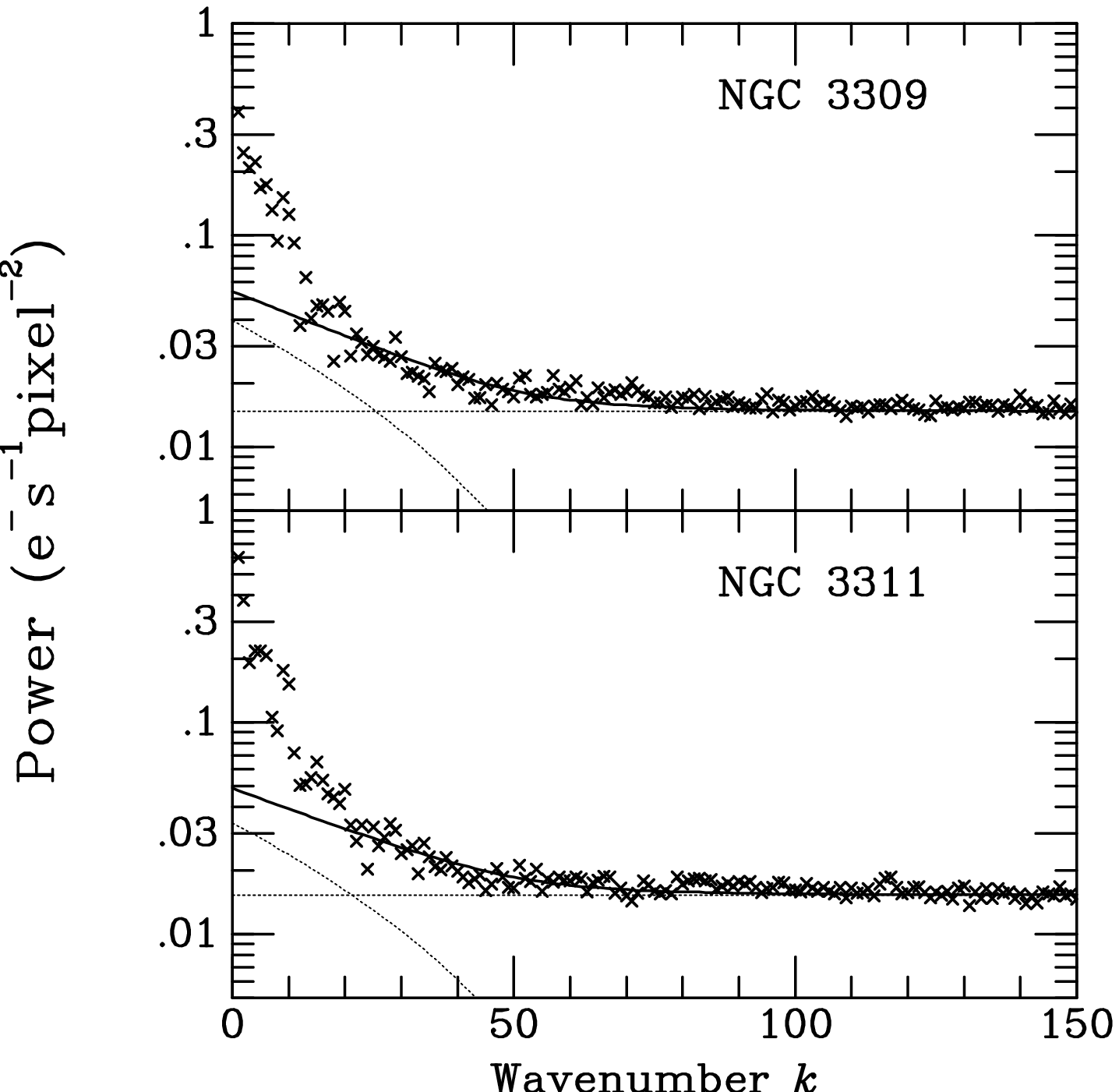}
\vspace{20pt}
\caption[Power Spectra for NGC~3309 and NGC~3311 in Hydra]
{Power spectra are plotted for NGC~3309 and NGC~3311 in the Hydra cluster.
Solid lines indicate the sum of $P_0{\times}E(k)\,{+}\,P_1$; the dotted lines
are the individual components.  The spectrum shown for NGC~3309 is for
all three annuli ($2\arcsec\,{<}\,r\,{<}\,48\arcsec$), 
while the spectrum for NGC~3311
excludes the inner annulus.
\label{hydrapowerspectra}}
\end{figure}

\begin{figure}
\figurenum{4}
\epsscale{0.65}
\plotone{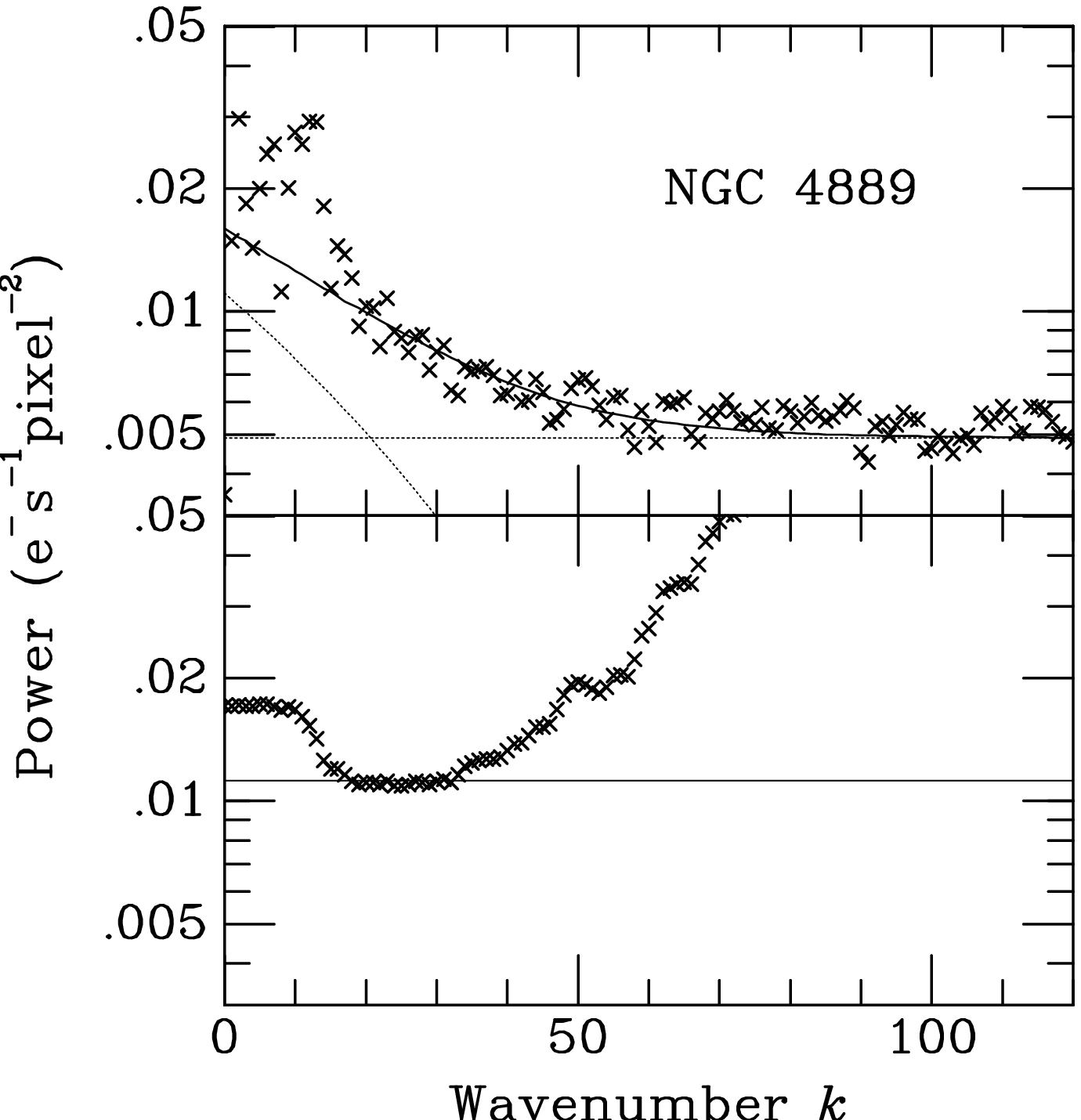}
\vspace{20pt}
\caption[Power Spectrum for NGC~4889 in Coma]
{The power spectrum for the annular region 
$12\arcsec\,{<}\,r\,{<}\,24\arcsec$ of NGC~4889 is plotted in the top panel.
The fitted lines are defined as in Figure \ref{hydrapowerspectra}.
The lower panel demonstrates the sensitivity of the $P_0$ fit to starting
wavenumber (i.e., ignoring wavenumbers lower than $k$ in the fit).  
We fitted the power spectrum in the range $20\,{<}\,k\,{<}\,33$
and the line in the lower plot indicates the adopted $P_0$
value.
\label{comapowerspectra}}
\end{figure}

To compute the fluctuation magnitude and distance to each galaxy, we first
measured the \Kp\ SBF power in three annular regions centered on each 
galaxy.   
Radii and $P_0$ powers measured are listed in Table~\ref{mbars}.  
The power spectra are plotted in 
Figures~\ref{hydrapowerspectra} and \ref{comapowerspectra}
with the individual components $P_0{\times}E(k)$ and $P_1$, and their sum.
We determined the fluctuation magnitudes \mKp\ corrected for 
residual variances using Equation~\ref{mkp}.  
Finally, the absolute fluctuation magnitude \MKp\ from JTL
was used to compute the distance moduli.

\subsection{Residual Variance from Globular Clusters and Galaxies}

JTL described the advantages of using optical images to 
identify and mask GCs and background galaxies in IR images
prior to performing the SBF analysis.  We applied this technique 
(as described in Section~\ref{howtosbf}) to the data presented in this paper.
We used the $I$-band image of NGC~4889 kindly provided by J. Blakeslee 
(1997, private communication) 
to identify GCs and galaxies down to approximately
$K\,{=}\,22.5$.  
A. Dressler (1997, private communication) 
generously made a deep $I$-band image of 
the Hydra galaxies available, allowing us to reach a similar limiting 
magnitude in our Hydra data.

\begin{figure}
\figurenum{5}
\epsscale{0.6}
\plotone{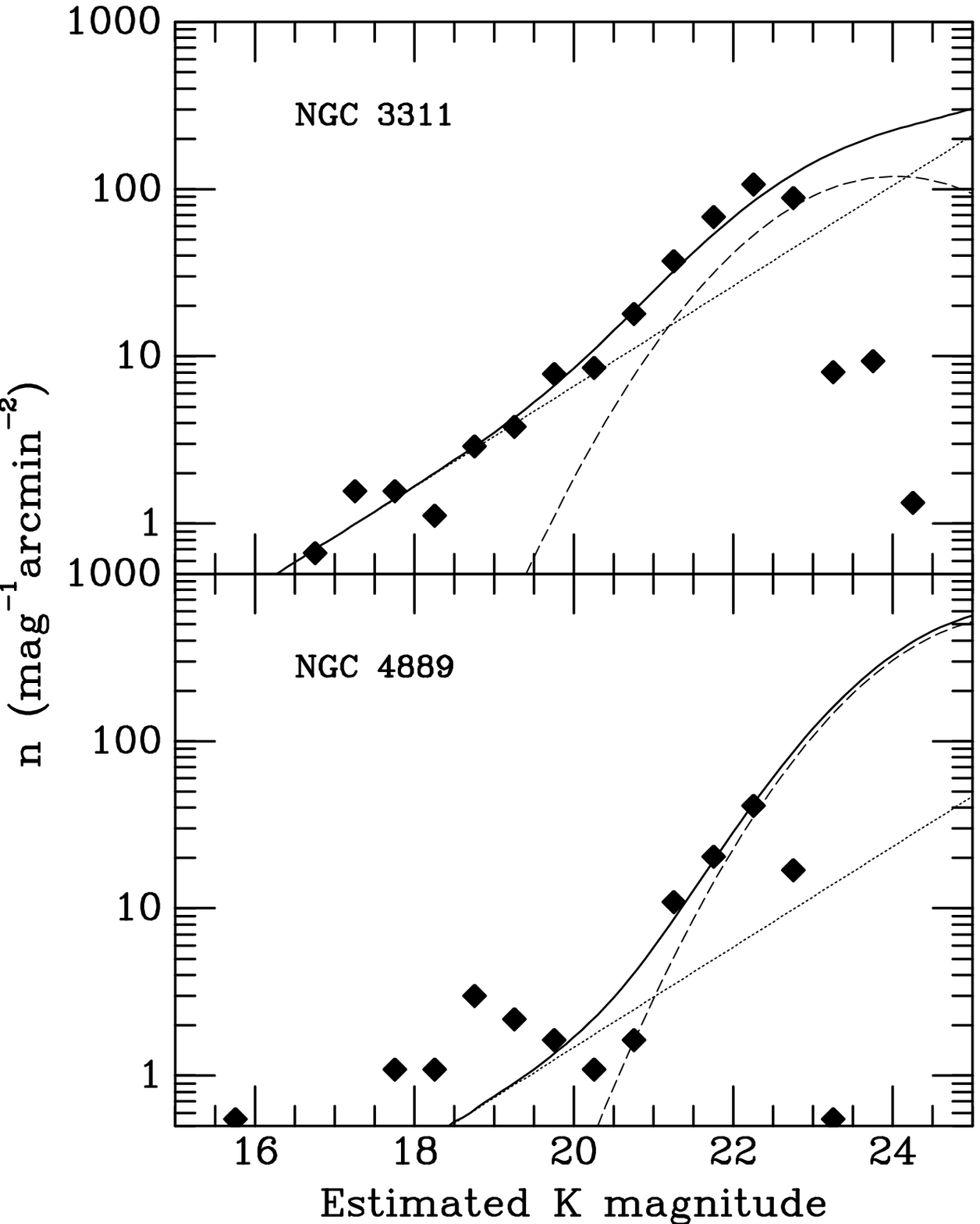}
\vspace{20pt}
\caption[GC and galaxy luminosity functions for Hydra and Coma]
{Luminosity functions are fitted to photometry of objects measured in the
$I$-band image and translated to $K$ by assuming uniform $(I{-}K)$ colors
for GCs and galaxies.  The extrapolated luminosity functions are used to
compute the variance from GCs and galaxies fainter than the completeness
limit, which is $K\,{\approx}\,22.5$ for these observations.  
The dotted and dashed lines show the individual fits to the galaxy and GC
luminosity functions, respectively, and the solid line is their sum.
The GCs in the Hydra image were measured as a single population 
centered on NGC~3311.
\label{gclf}}
\end{figure}

To constrain the GCLF at \Kp,
we adopted uniform GC and galaxy colors and translated $I$ magnitudes
to $K$ as described in the previous section.
We then performed the fit to the GC and galaxy luminosity
functions as if the data had been collected from a $K$-band image and
created a mask of all objects brighter than the completeness limit.
The luminosity functions were integrated beyond the cutoff magnitude
as a function of radius from the center of each galaxy, and we computed
the residual variance $P_r$ for each annular region.  $P_r$ values
are listed in Table~\ref{mbars}.  The luminosity functions are plotted
in Figure~\ref{gclf}.

We adapted the models developed by Blakeslee \& Tonry (1995)
to check our estimate of
the residual variance contributed by 
GCs and galaxies.  
If the GC specific frequency is $S_N\,{=}\,7$
for NGC~4889 (Blakeslee \& Tonry 1995) 
then ${\sigma^2_{GC}\,{/}\,\sigma^2_{SBF}}\,{\approx}\,0.6$, 
which corresponds to a correction of $P_r\,{\approx}\,0.4\,P_0$.
We note, however, that the specific frequency measured by Blakeslee \&
Tonry is smaller in the inner regions of NGC~4889 by a factor of two to three.
The variance is proportional to $S_N$, so we expect to find 
$P_r\,{\approx}\,0.2\,P_0$ within 48\arcsec\ of the center.
Examination of the $P_r/P_0$ values listed in Table~\ref{mbars} confirms
that the correction for undetected GCs and galaxies is less than 20\%
for NGC~4889.  
This assumes no additional $P_g$ correction for residual 
spatial variance in the background.
If $P_g$ is subtracted from $P_0$ prior to making the comparison to $P_r$, 
we find that $P_r$ is 0.2 to 0.4 times the fluctuation power, just
as the models predict.

NGC~3311 in Hydra has a high specific frequency
$S_N\,{=}\,15$ (McLaughlin et al. 1995), 
but the Hydra cluster is much closer than Coma and the optical images are 
comparably deep, so the GCLF for NGC~3311 is better sampled 
(in fitting the GCLF, 
we assumed that all GCs detected belong to the NGC~3311 system).  
The $P_r$ correction for the Hydra galaxies is less 
than 2\% for most annuli. 
Subtracting  $P_g$ from $P_0$ results
in a GC and galaxy contribution to the fluctuation power between
3 and 6\%.  The model predicts a $P_r$ correction of $0.25\,P_0$ 
for NGC~3311.  It is likely that $S_N$ is lower near the center of
NGC~3311 where the galaxy is brightest.

It is encouraging how well the point-source masks derived from optical
data reduce the residual variance from undetected GCs and galaxies.  
Even though several assumptions about the GCLF must be made to
compute $P_r$, the fluctuation magnitude is not sensitive to the exact
form of the luminosity function.  
For example, if the distance
used to fix the peak of the GCLF for NGC~4889 is changed
from 70 to 110 Mpc, the result is a change of only 0.03 mag in \mKp.
Similarly, changing the width of the GCLF, or the 
relative fractions of galaxies and GCs, or the bright cutoff magnitude,
or any of the other input parameters, had an even smaller effect 
on the final fluctuation magnitude.  These details may be important for
measuring a distance based on the absolute magnitude of the GCLF peak,
but they do not significantly affect the \Kp\ fluctuation magnitude.
There is sufficient covariance between the fit parameters that if the
fit to either the GC or galaxy luminosity function is in error, the fitting
procedure will compensate by adjusting the other,
resulting in a fit that comes close to correctly predicting $P_r$. 

\begin{deluxetable}{cccccccc}
\tablecaption{\Kp\ SBF Magnitudes in Hydra and Coma\label{mbars}}
\tablewidth{0pc}
\tablehead{
\colhead{Galaxy}&
\colhead{Radius (\arcsec)} &
\colhead{$P_0$ (\epspp)} &
\colhead{$P_0/P_1$} &
\colhead{$P_r/P_0$}&
\colhead{$P_g/P_0$}&
\colhead{\snsbf}&
\colhead{\mKp (mag)}}
\startdata
NGC~3309 & 2--12 & $1.62\pm0.14\times10^{-2}$ & 7.2 & 0.00 & 0.29 & 1.65 & $27.72 \pm 0.11$ \nl
         &12--24 & $2.88\pm0.10\times10^{-2}$ & 3.5 & 0.01 & 0.60 & 0.44 & $27.74 \pm 0.13$ \nl
         &24--48 & $7.42\pm0.39\times10^{-2}$ & 2.3 & 0.01 & 0.84 & 0.12 & $27.72 \pm 0.42$ \nl
         & 2--48 & $3.95\pm0.14\times10^{-2}$ & 2.7 & 0.01 & 0.72 & 0.25 & $27.78 \pm 0.20$ \nl
\tablevspace{10pt}
NGC~3311 & 2--12 & $1.09\pm0.21\times10^{-2}$ & 3.1 & 0.04 & 0.42 & 0.74 & $28.42 \pm 0.22$ \nl
         &12--24 & $2.34\pm0.10\times10^{-2}$ & 2.7 & 0.02 & 0.48 & 0.58 & $27.69 \pm 0.13$ \nl
         &24--48 & $3.91\pm0.23\times10^{-2}$ & 2.1 & 0.02 & 0.67 & 0.27 & $27.66 \pm 0.27$ \nl
         &12--48 & $3.35\pm0.16\times10^{-2}$ & 2.2 & 0.02 & 0.64 & 0.31 & $27.72 \pm 0.23$ \nl
\tablevspace{10pt}
NGC~4889 & 2--12 & $0.83\pm0.12\times10^{-2}$ & 6.0 & 0.15 & 0.19 & 1.90 & $28.55 \pm 0.16$ \nl
         &12--24 & $1.11\pm0.02\times10^{-2}$ & 2.3 & 0.15 & 0.53 & 0.33 & $29.03 \pm 0.13$ \nl
         &24--48 & $2.53\pm0.12\times10^{-2}$ & 1.6 & 0.11 & 0.76 & 0.09 & $29.09 \pm 0.33$ \nl
         & 2--24 & $0.86\pm0.05\times10^{-2}$ & 2.9 & 0.18 & 0.41 & 0.55 & $28.99 \pm 0.11$ \nl
         &12--48 & $1.88\pm0.08\times10^{-2}$ & 1.8 & 0.12 & 0.69 & 0.15 & $29.05 \pm 0.24$ \nl
\enddata
\end{deluxetable}

\subsection{Additional Sources of Residual Variance}

\begin{figure}
\figurenum{6}
\epsscale{0.65}
\plotone{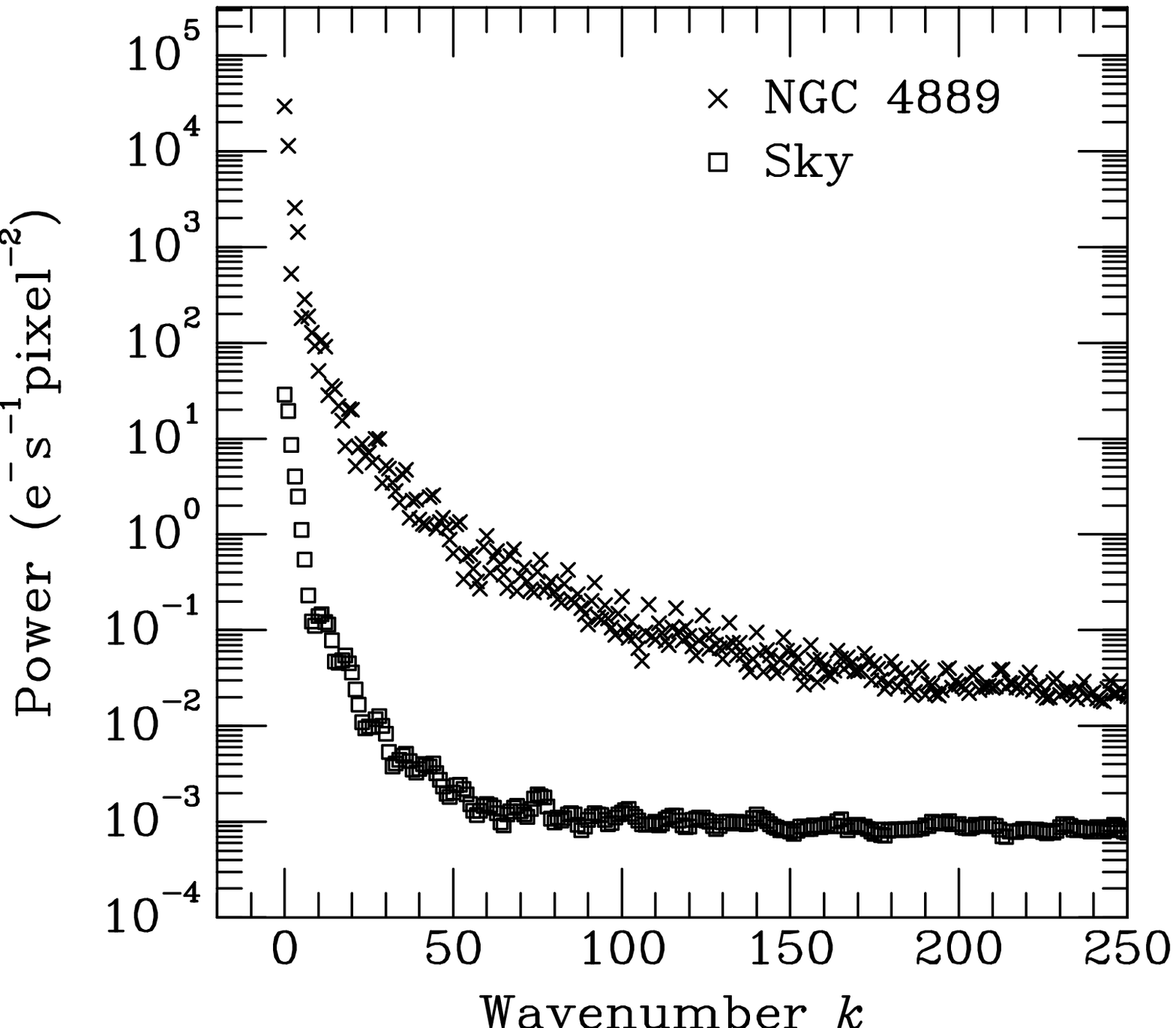}
\vspace{20pt}
\caption[Power Spectra for NGC~4889 and an Adjacent Sky Region]
{The power spectrum for NGC~4889 
is shown again, but unscaled by the mean galaxy brightness as 
in Figure~\ref{comapowerspectra}.
The figure also shows the power spectrum taken from an adjacent ``sky''
region approximately 1\arcmin\ north of the center of NGC~4889.  
The deviation of the sky spectrum from a flat
white-noise power spectrum is evident, particularly for 
$20\,{<}\,k\,{<}\,50$ where the fits are typically performed.
Some of the excess power is from stellar fluctuations in the 
outer regions of NGC~4889. 
Subtracting the sky
power spectrum from the data spectrum gives the variances
listed in Table~\ref{skypowerspectra}, which are 
consistent with the \mKp\ magnitudes calculated by assuming a uniform 
residual variance $P_g$ proportional to $P_1$.  
The consistency of our final \mKp\ values between annuli suggests that we are 
correctly removing the extra power present
in the background sky spectrum for $20\,{<}\,k\,{<}\,50$.
\label{skypower}}
\end{figure}

After correcting for undetected GCs and galaxies, we still found
that $P_0$ increased significantly with radius as the $S/N$ decreased.  
Jensen et al. (1996) 
demonstrated that low-$S/N$ IR SBF
observations can be biased by the presence of residual spatial patterns 
resulting from dark current variations, flat fielding, and 
subtraction or masking of faint, 
diffuse objects in the presence of a noisy background.
Bright objects leave residual dark current 
in the HAWAII array that decay with time,
adding to the background noise on scales corresponding to the
spacing between dithered observations.
The noise in our data shows an increase in power at
low wavenumbers ($k\,{<}\,50$) above the fitted constant white noise $P_1$
(Figure~\ref{skypower}).
We estimated that the maximum radial change in \mKp\ within 50\arcsec\ of
the galaxy center due to hypothetical 
stellar population variations is ${\lesssim}0.5$ mag.
The radial gradients in $P_0$ on the order of a factor of three
must therefore be the result of residual
variance that is proportional to the area of the annulus.  

To determine the importance of residual variance in our images, we measured
the power spectrum of four relatively empty regions 
of the Coma image as far from the
center of the galaxy as the field of view would permit.  
GCs and background galaxies were masked as before. 
The power spectrum for the sky regions do contain some SBF signal from the 
galaxy, but background noise dominates the stellar fluctuations at these radii.
The power spectrum of one of the regions is 
shown in Figure~\ref{skypower} and is clearly
not flat, with increasing power towards $k\,{=}\,0$.
Comparison of the power spectra from various regions in the image show
that the deviation from a constant noise spectrum ($P_1$ constant with $k$)
is roughly proportional to $P_1$.
The additional noise in the region of the spectrum where the fits are
performed ($20\,{<}\,k\,{<}\,50$) can be a significant fraction of $P_0$.

To remove the background spatial variance, 
we subtracted the uniform correction $P_g$ 
that best mitigates the radial gradient in $P_0$.
Because the residual variance is assumed to be uniform across the 
field of view, $P_g$ is proportional to the area of the region being analyzed.
$P_1$ also increases with the area of the annulus, and $P_g$ is roughly
proportional to $P_1$.  
Adopting a uniform residual variance correction is a simple way to deal
effectively with the power in the spectrum contributed by residual
patterns in the image. 
The levels of $P_g$ required to correct for the residual variance in our data
were quite large, often 50\% or more of the total power measured.
In terms of the sky brightness, the $P_g$ corrections were comparable to those
measured 
by JTL:  
for NGC~3309 $P_g$ was 0.117\% of the sky brightness;
NGC~3311 was best corrected with $P_g\,{=}\,0.10\%$ of the sky level; for
NGC~4889 in Coma we applied a correction of 0.07\% of the sky brightness.
These corrections are large compared to $P_0$ for the distant galaxies 
in this study, but they are comparable to the corrections applied 
by JTL that were typically 0.1\% of the background level.

\begin{deluxetable}{cccccc}
\tablecaption{NGC~4889 Fluctuation Amplitudes with Sky Power Spectra 
Subtracted
\label{skypowerspectra}}
\tablewidth{0pc}
\tablehead{
\colhead{Radius} &
\multicolumn{4}{c}{Power (${\times}10^{-2}\,\epspp$)} &
\colhead{Estimated}\nl
\colhead{(\arcsec)} &
\colhead{Sky 1} &
\colhead{Sky 2}&
\colhead{Sky 3}&
\colhead{Sky 4}&
\colhead{uncertainty}}
\startdata
 2--12 & 0.48 & 0.60 & 0.51 & 0.29 & 0.09   \nl
12--24 & $<0$ & 0.49 & 0.43 & $<0$ & 0.13   \nl
 2--24 & 0.20 & 0.54 & 0.45 & $<0$ & 0.13   \nl
\enddata
\end{deluxetable}

Subtracting a uniform residual variance $P_g$ does a reasonably good job
of removing the radial gradient in $P_0$, but does it correctly remove
the noise observed in the power spectrum of the sky region?
To answer this question, we took the power spectra measured in the four
surrounding regions and subtracted them individually from the 
data power spectrum 
for NGC~4889 prior to measuring $P_0$.  
The sky power spectra were measured in apertures with the same radii
as those used to measure $P_0$ in NGC~4889.
The power spectrum from one region is shown in Figure \ref{skypower},
along with the unscaled power spectrum from NGC~4889 for reference 
(in Figure \ref{comapowerspectra} the power spectrum was normalized by the
mean galaxy brightness).
Table~\ref{skypowerspectra} lists the fluctuation powers
measured in the inner two regions after subtracting the power spectrum from the
specified sky region,
along with an estimate of the uncertainty for each annulus.   
The variances listed in Table~\ref{skypowerspectra}
are consistent with the values determined by applying the $P_g$ correction;
however, it is clear that variations between power spectra from 
different regions of the image significantly affect the \mKp\ measured. 
In some cases the sky power spectrum had so much power at low wavenumbers
that subtracting it from the galaxy power spectrum 
resulted in a negative variance.
We attempted to pick typical areas of the image in which to measure
the sky power spectrum.  However, if the region near the center of 
NGC~4889 is as grubby as some of the lower-$S/N$
outer regions of the frame (which
it should not be), then NGC~4889 could be more distant than we have
estimated by perhaps ${\sim}0.5$ mag. 

The variances listed in Table~\ref{skypowerspectra} confirm that the
method of correcting for residual variance by subtracting $P_g$ is
a reasonable way to account for power in the sky spectrum.
While this experiment cannot rule out the possibility that the majority
of the power measured is the result of structure in the noise, 
it constrains the range of fluctuation magnitudes.  According to the
data listed in Table~\ref{skypowerspectra}, \mKp\ must be brighter than
${\sim}29.6$ (corresponding to a power of $0.2{\times}10^{-2}\ \epspp$),
since sky power spectra that subtracted enough
power in the fit region to get this fluctuation magnitude 
also over-subtracted other annuli.
On the other hand, \mKp\ cannot be any brighter than ${\sim}28.5$ mag
($0.6{\times}10^{-2}\ \epspp$).
The fluctuation magnitudes we compute for NGC~4889 
by subtracting a uniform background spatial
variance $P_g$ to minimize the radial change in $P_0$ fall comfortably
within this range (Table~\ref{mbars}).
The sky power subtraction experiment was
not repeated for the Hydra data because the galaxies almost completely fill
the field of view.

Assuming uniform colors for the GC and galaxy populations to determine
the $P_r$ correction potentially
introduces a bias in the fluctuation magnitudes we measure.
Some faint galaxies are extremely red, and if they are not 
masked (because they are fainter than the completeness limit in $I$), 
they will add more flux to the SBF power spectrum at \Kp\ 
than we subtract with $P_r$.
Approximately four objects were easily detected in the \Kp\ image 
of NGC~4889 that were below the completeness limit in the $I$-band image.  
These were 
masked prior to computing $P_0$, but additional faint red galaxies 
certainly lurk in the noise.
Thus the $P_r$ correction is probably underestimated.
At this point we cannot say how much additional power must be subtracted
from $P_0$ to account for very red objects in our field without knowing
the color distributions of GCs and background galaxies.  
At an approximate level, though, we account for the deficit in $P_r$
with our uniform residual variance correction $P_g$.  If the extremely
red objects (mostly background galaxies) are distributed uniformly
in the field of view, then they will most strongly influence the outer
annuli, causing a radial increase in $P_0$. 
By adopting a level of residual variance $P_g$ that best removes the
radial trend in $P_0$, we simultaneously 
correct for both residual patterns and
uniformly distributed red galaxies.
We suggest that in future observations,  
optical integrations be deep enough to reduce the $P_r$
correction to negligible levels, thereby reducing the systematic error
associated with $P_r$.  
For this paper, we increased the estimated uncertainty 
for $P_r$ listed in Table~\ref{errorbudget} 
to account for the potential bias.  We are confident that the
bias is largely removed by our residual 
variance subtraction.  

\subsection{Reliability of the SBF Measurements}

The stellar SBF $S/N$ (with residual variances subtracted) 
was computed for each annulus 
using the definition from JTL:
\begin{equation}
\snsbf = {{P_0-P_r-P_g} \over {P_1+P_g}}
\label{snrsbf}
\end{equation}
and listed in Table~\ref{mbars}.
In calibrating the \Kp\ SBF scale, 
JTL 
showed that measurements with
$\snsbf\,{<}\,1$ are unreliable.  Only the innermost regions of the galaxies
in the current study have strong enough signals to meet this criterion.
Smaller annuli are also the least sensitive 
to the residual variance correction.
After the residual variance bias has been removed, we find that the
fluctuation magnitudes are quite consistent, in spite of the low
$S/N$ in the outer apertures.  

Fluctuation magnitudes measured in the innermost annulus (2\arcsec--12\arcsec)
of NGC~3311 and NGC~4889 fail to match the outer annuli.
The inner region of NGC~3311 has an anomalously faint \mKp\ magnitude.  
The center of this galaxy has an unusual off-center nucleus,
and the galaxy fitting routine fails near the center.
We masked the center of NGC~3311 prior to measuring the SBFs, but it is
not clear if the anomalous fluctuation magnitude is the result of a
distinct stellar population in the nuclear region of NGC~3311 or due to
a failure in the SBF analysis.  
It is also possible that the nuclear region of NGC~3311 is obscured
by dust.  A uniform distribution of dust will reduce the measured
fluctuation amplitude.
The inner annulus is excluded from the
power spectrum plotted in Figure~\ref{hydrapowerspectra} and from the
calculation of \mKp\ for this galaxy.
SBFs in the central region of NGC~4889 are brighter than in the outer
annuli after correction for residual variance.  Both $I$ and \Kp\ images
show a similar pattern near the center that may be blended
images of GCs or stellar SBFs.
They may also signal the failure of our galaxy model fitting routine,
which sometimes has difficulty very near the centers of galaxies.
Since the fluctuation magnitude is 0.5 mag brighter in the inner annulus
than in any of the other annuli (including composite regions that 
include the inner annulus but weight it less heavily), it seems likely
that the inner measurement is biased by point sources or a deviation of
our model from the real surface brightness profile near the center. 
The power spectrum shown in Figure~\ref{comapowerspectra} is for the
second annulus ($12\arcsec\,{<}\,r\,{<}\,24\arcsec$). 

Uncertainties in many different input parameters affect the fluctuation 
magnitudes measured.
To estimate the uncertainty in \mKp, we changed the input values for
each parameter in turn and measured the change in \mKp.  The uncertainties
from PSF fitting, galaxy subtraction, residual background subtraction,
correction for GCs, and so forth were added in quadrature.
This procedure assumes that the uncertainties are independent, 
which is not strictly
true; significant covariances exist between sky level subtraction and
galaxy profile fitting, for example, or between the photometric zero point
and the PSF normalization.
The individual uncertainties for the Hydra and Coma observations are 
specified in detail in Table~\ref{errorbudget}.  
Some of these uncertainties are systematic to our study, so the 
uncertainties in our final fluctuation magnitudes are larger than the 
scatter between individual measurements.
The uncertainty in the photometric zero point, for example, is the
same for both galaxies in the Hydra image.  
The uncertainties in fluctuation magnitudes measured in different annuli 
have a large covariance as well. 

Other sources of uncertainty could be either random or systematic, or both.
The estimated uncertainty in the 
correction for any residual spatial variance ($P_g$) is
the dominant source of uncertainty listed in Table~\ref{errorbudget}.
Whether or not this uncertainty affects all annuli equally (systematically)
or not depends on the distribution of the background variance in the image.
The uncertainty in $P_g$ is particularly difficult to estimate
because many sources of variance add to the power measured
(residual dark current from saturated sources, unmasked objects in sky
images, etc.).
We corrected for background spatial variance as well as possible
and estimated realistic uncertainties, but adding
systematic uncertainties 
in quadrature leads to an underestimate of the uncertainty
in \mKp.  To account for the fact that the uncertainty in $P_g$ has both 
random and systematic components, we estimated an additional 
systematic uncertainty 
from residual background variance subtraction and included it
in Table~\ref{errorbudget}.  The correction for residual variance
from undetected GCs and galaxies also has random and systematic
uncertainties associated with it, and we added an additional systematic
uncertainty in $P_r$ to Table~\ref{errorbudget}.
The estimated 
systematic uncertainties were not included in the \mKp\ uncertainties,
but were added in quadrature to compute the uncertainties in the
distance moduli, distances, and Hubble constant.

Although it is difficult to understand all the possible covariances in
the estimated uncertainties, we feel that we have fairly estimated the 
uncertainty in our measurement of \mKp.  
The uncertainties reported by JTL 
were determined by adding the individual uncertainties in quadrature, 
and the fits to that data 
had $\chi^2$ values per degree of freedom between 0.5 and 2, 
suggesting that the estimated uncertainties were reasonable.
In the final analysis, our fluctuation magnitudes are consistent, 
indicating that residual variance has been removed properly in 
annuli of different sizes.

\begin{deluxetable}{lcc}
\tablecaption{SBF Uncertainties for Hydra and Coma \label{errorbudget}}
\tablewidth{0pc}
\tablehead{
\colhead{Source of} &
\colhead{Hydra}&
\colhead{Coma}\nl
\colhead{Uncertainty}&
\colhead{$\sigma$ (mag)} &
\colhead{$\sigma$ (mag)} }
\startdata
{\it ~~~~Random Uncertainties} & & \nl
Photometry\dotfill& 0.01 & 0.01 \nl
PSF normalization\dotfill& 0.02 & 0.03 \nl
$P_0$ fit (typical)\dotfill& 0.04 & 0.05 \nl
Residual sky and & & \nl
~~~~~model galaxy subtraction\dotfill& 0.05 & 0.03 \nl
GCs and galaxies\dotfill& 0.03 & 0.05 \nl
Residual noise subtraction\dotfill& 0.03--0.41 & 0.02--0.32 \nl
&&\nl
{\it ~~~Systematic Uncertainties} & & \nl
GCs and galaxies\dotfill& 0.05 & 0.10 \nl
Residual noise subtraction\dotfill& 0.16 & 0.15 \nl
\Kp\ SBF Calibration\dotfill& 0.06 & 0.06 \nl
Cepheid zero point\dotfill& 0.10 & 0.10 \nl
\enddata
\end{deluxetable}

\subsection{Distances}

To compute the distance moduli, we used the \Kp\ SBF calibration from
JTL based on Cepheid distances to M31 and the Virgo cluster.
JTL found that
$\MKp\,{=}\,-5.61\,{\pm}\,0.12$, 
in excellent agreement with Jensen et al. (1996),
Luppino \& Tonry (1993), and Pahre \& Mould (1994).
No significant trend in \MKp\ was found as a function of galaxy
color for the limited range $1.15\,{<}\,\v-i\,{<}\,1.27$.  
There are no published \v-i\ colors for the galaxies in this sample.
To confirm that the calibration applies to the Coma and Hydra cluster
galaxies, we estimated \v-i\ colors using published $(V{-}K)$
values from (Persson, Frogel, \& Aaronson 1979) and mean $(I{-}K)$
colors for our calibration galaxies 
(JTL).
For the Virgo, Fornax, and Eridanus galaxies with $\v-i\,{>}\,1.2$, 
the mean $(I{-}K)\,{=}\,2.2$,
implying $\v-i\,{\approx}\,1.23$ for NGC~4889.
The estimated \v-i color for NGC~4889 is typical of 
the JTL sample,
and we adopt the calibration from JTL without modification.
The \v-i colors were estimated to be 1.25 and 1.31 for NGC~3309 and NGC~3311,
respectively.  NGC~3311 may be redder than the galaxies used 
by JTL 
to calibrate the \Kp\ SBF distance scale, but the uncertainty on the
estimated \v-i color is ${\sim}0.1$ mag so 
we cannot conclude that NGC~3311 is different from the 
galaxies in the JTL sample.
The same calibration of \MKp\ was used for the Hydra galaxies.
Because the K corrections predicted by old stellar population models are
negligible at $K$ (JTL), we computed the luminosity distance (in Mpc)
using the relation
\begin{equation}
d_L = 10^{\,0.2\,({\overline m}\,{-}\,{\overline M})\,{-}\,5}
\label{d_L}
\end{equation}
and found a distance of $84.7\pm 9.8$ Mpc to NGC~4889.  
The distances to the Hydra cluster galaxies were computed in the same way
giving distances of $46.3\pm4.9$ and $45.7\pm5.0$ Mpc to 
NGC~3309 and NGC~3311.
The distances are remarkably consistent (the uncertainties for NGC~3309 and
NGC~3311 are correlated because both measurements were made in the 
same image and share many of the same uncertainties).  
Fluctuation magnitudes, distance moduli and distances are listed in 
Table~\ref{distances}.  The distance uncertainties include the 
systematic uncertainties from the calibration, added in quadrature.

The astrometric satellite Hipparcos has recently provided important 
parallax distances to Cepheid variable stars in the Galaxy.  
Hipparcos measurements suggest that the Cepheid
distance scale should be increased by ${\sim}10\%$ (Feast \& Catchpole 1997).  
If confirmed, the Hipparcos results may require us to modify the Cepheid
calibration of IR SBFs.
Madore \& Freedman (1997), however, find that the change in Cepheid
zero point suggested by the Hipparcos data is smaller than the 
uncertainty in the Cepheid calibration due to uncertainties in 
metallicity or reddening corrections,
so it is premature to adopt a new calibration at this point.

\section{The Hubble Constant\label{h0}}

We computed the Hubble constant using the equation
\begin{equation}
H_0 = {{cz}\over{d_L}}.
\label{easyH0}
\end{equation}
At the distance of the Coma cluster ($z\,{=}\,0.025$), 
the cosmological correction to the luminosity distance $d_L$ is 
less than 1\% for $q_0\,{=}\,0.5$ (Kolb \& Turner 1990, Eq. 2.52).
We used the distances and velocities listed in 
Table~\ref{distances} to compute the tabulated values of $H_0$.
Mean cluster velocities in the CMB frame were drawn from Han \& Mould (1992).
The cluster velocities reported by Faber et al. (1989) are virtually
identical (4033 \kms\ for Hydra and 7202 \kms\ for Coma).
Values of the Hubble constant for
the Hydra and Coma clusters are consistent (Table~\ref{distances}), 
giving a mean Hubble constant of $H_0\,{=}\,87\,{\pm}\,11$ \kmsmpc.
The uncertainties for the three measurements are not completely 
independent because they are drawn from the same observations.

\begin{deluxetable}{cccccc}
\tablecaption{\Kp\ SBF Distances to Hydra and Coma\label{distances}}
\tablewidth{0pc}
\tablehead{
\colhead{Galaxy}&
\colhead{\mKp} &
\colhead{$(\overline m{-}\overline M)_{\Kp}$} &
\colhead{$d$}&
\colhead{$v$\tablenotemark{a}}&
\colhead{$H_0$}\nl
&\colhead{(mag)}&
\colhead{(mag)}&
\colhead{(Mpc)}&
\colhead{(\kms)}&
\colhead{(\kmsmpc)}}
\startdata
NGC~3309 & $27.72\pm0.11$ & $33.33\pm0.23$ & $46.3\pm4.9$ & $4054\pm296$ & $87\pm11$ \nl
NGC~3311 & $27.69\pm0.13$ & $33.30\pm0.24$ & $45.7\pm5.0$ & $4054\pm296$ & $89\pm12$ \nl
NGC~4889 & $29.03\pm0.13$ & $34.64\pm0.25$ & $84.7\pm9.8$ & $7186\pm428$ & $85\pm11$ \nl
\enddata
\tablenotetext{a}{Han \& Mould 1992}
\end{deluxetable}

The IR SBF Hubble constant measured in this study is consistent with 
an extensive optical ($I$-band)
SBF survey of several hundred galaxies that
gives $H_0\,{=}\,81\,{\pm}\,6$ \kmsmpc\ (Tonry et al. 1997).
A detailed comparison of the $I$-band SBF distance scale with 
several other distance estimators is included in Tonry et al. (1997).
Lauer et al. (1997) and Ajhar et al. (1997) used the Hubble Space Telescope
(HST) to measure $I$-band
SBF distances to several Abell clusters.
Calibrating the brightest cluster
galaxy distance scale using $I$-band SBFs gives
$H_0\,{=}\,89\,{\pm}\,10$ \kmsmpc.

The distance we measured to NGC~4889 is shorter than the distance of
$102\pm14$ Mpc recently measured
by Thomsen et al. (1997) using the HST to probe $I$-band SBFs in  
the Coma cluster galaxy NGC~4881.
The two distances agree at the 1-$\sigma$ level;
nevertheless, we do not know if the 
difference between our measurement and theirs is significant.
Both our measurement and that of Thomsen et al. have low 
$S/N$ (Thomsen et al. accumulated only 0.7 SBF photons
per noise photon).
The detectors, filters, and procedures used by Thomsen et al. are 
significantly different from those used for this study, making it difficult
to compare possible sources of systematic error. 
Previous IR SBF studies have found
a few galaxies with anomalously bright fluctuation magnitudes
(Luppino \& Tonry 1993; Pahre \& Mould 1994;
Jensen et al. 1996; JTL).
Some were shown to be anomalous
because of observational bias at low $S/N$
(JTL); 
others may have unusual stellar populations and await 
follow-up observations to confirm their bright fluctuation magnitudes.
It is possible that NGC~4889 in Coma is a galaxy with anomalously 
bright IR SBFs because it possesses an unusual stellar population.
On the other hand, 
there is no obvious reason why NGC~4889 should have a different stellar
population from the other galaxies we have studied, and we believe our
distance measurement to be reliable.
Indeed, we have measured IR SBFs in three galaxies in two clusters
with remarkably consistent results 
(although the agreement between NGC~3309 and NGC~3311 is partially a
result of the fact that they were observed simultaneously).

These are the first IR SBF measurements to be made at such large
distances, and additional high-$S/N$ observations will be needed to 
reduce the uncertainty in the SBF distance to Coma.
The HST with NICMOS and 
new large telescopes equipped with large-format IR detectors will allow
us to confirm our results using larger samples.
Improvements in observational and data reduction techniques will
reduce residual variances and increase the reliability of IR SBF
measurements.
In future ground-based 
observations we will use deep optical images to identify faint
objects in sky fields as we did in the galaxy images to improve  
sky subtraction and reduce background patterns from unmasked 
objects in the sky fields.  The influence of residual patterns 
from bright saturated objects will be reduced by using shorter exposure
times and by dithering in only one direction.  By dithering along one
axis only, the power from residual patterns will be confined
to a line and can be masked prior to fitting the power spectrum.
These modifications to the observing techniques should significantly
reduce the amplitude of residual background patterns that can bias
low-$S/N$ IR SBF measurements.

\section{Summary}

This paper describes our attempt to extend the IR SBF distance
scale to the Coma cluster.  We measured \Kp\ SBFs in
two Hydra cluster giant elliptical galaxies and in NGC~4889 in Coma.    
Our conclusions are summarized as follows:

1.  IR SBF distance measurements to three galaxies in the Hydra and 
Coma clusters imply a Hubble constant $H_0\,{=}\,87\,{\pm}\,11$ \kmsmpc\  
if sources of residual variance have been subtracted correctly.  
Results from both clusters are consistent.  
These measurements support the conclusion
that Hydra's radial peculiar velocity is small.

2.  Fluctuations from globular clusters and background galaxies can be 
adequately removed, even at distances of 7000 \kms.  
Modest optical exposures can be used to identify and mask GCs and galaxies
in IR images, potentially reducing the variance from undetected GCs and
galaxies to negligible levels.
Without using optical images, considerably longer IR integrations could be 
required to reduce the GC correction to similarly low levels.

3.  Corrections to remove residual spatial variance in the background
 are necessary to 
measure radially consistent fluctuation magnitudes.  
The power spectrum of the noise in our near-IR images is not 
independent of wavenumber.  
For the relatively short integration on NGC~4889 described in this study,
the size of the correction for residual variance is large compared to the 
SBF signal.  
To achieve reliable results in the future,
IR SBF researchers should strive
to produce images that are as clean and flat as possible, and to integrate
long enough under good seeing conditions to achieve a high $S/N$.
Additional improvements in observing techniques may also help 
reduce residual background variance in future IR SBF measurements.

The SBF observations described in this study were obtained using relatively
short integration times on a modest 2.24 m telescope and using the first
of a new generation of 1024$^2$-pixel IR detectors. 
Future IR SBF studies of galaxies to 100 Mpc and beyond will benefit from 
NICMOS on the Hubble Space Telescope with its low background, and from the
new generation of 8~m class telescopes with clean, large-format IR detectors.
For example, the Gemini telescope on Mauna Kea with the NIRI infrared imager
under good seeing conditions will make it possible to make high-$S/N$
$K$-band SBF measurements at the distance of the Coma cluster in 
${\lesssim}2000$~s per galaxy.  With telescopes like Gemini we will be able to 
reliably measure SBF distances to many galaxies in a distant cluster 
in a night, making it possible to measure the Hubble constant,
map peculiar velocities, and measure
bulk motions on scales much larger than currently possible using SBFs.

\acknowledgments
We are indebted to John Blakeslee and Alan Dressler who provided deep
optical images of NGC~3309, NGC~3311, and NGC~4889.  
Their images were invaluable in reducing
the GC and galaxy contributions to the SBF signal to manageable levels.
This research was supported in part by grants STScI \mbox{GO-06579.01-95A}
and NSF AST9401519.


\end{document}